# Ultra-broadband polarization beam splitter and rotator based on 3D-printed waveguides

A. Nesic[1†*], M. Blaicher[1,2†], P. Marin-Palomo[1], C. Füllner[1], S. Randel[1], W. Freude[1], C. Koos[1,2,3*]

[1]Institute of Photonics and Quantum Electronics (IPQ), Karlsruhe Institute of Technology (KIT), Karlsruhe, Germany
[2]Institute of Microstructure Technology (IMT), Karlsruhe Institute of Technology (KIT), Eggenstein-Leopoldshafen, Germany
[3]Vanguard Automation GmbH, Karlsruhe, Germany
[†]These authors contributed equally to this work.
*e-mail: aleksandar.nesic@kit.edu, christian.koos@kit.edu

**Abstract:** Multi-photon lithography[1–3] has emerged as a powerful tool for photonic integration, allowing to complement planar photonic circuits by 3D-printed freeform structures such as waveguides[4,5] or micro-optical elements[6,7]. These structures can be fabricated with high precision on the facets of optical devices and lend themselves to highly efficient package-level chip-chip-connections in photonic assemblies[5]. However, plain light transport and efficient coupling is far from exploiting the full geometrical design freedom that is offered by 3D laser lithography. Here, we extend the functionality of 3D-printed optical structures to manipulation of optical polarization states. We demonstrate compact ultra-broadband polarization beam splitters (PBS) that can be combined with polarization rotators (PR) and mode-field adapters into a monolithic 3D-printed structure, fabricated directly on the facets of optical devices. In a proof-of-concept experiment, we demonstrate measured polarization extinction ratios beyond 11 dB over a bandwidth of 350 nm at near-infrared (NIR) telecommunication wavelengths around 1550 nm. We demonstrate the viability of the device by receiving a 640 Gbit/s dual-polarization data signal using 16-state quadrature amplitude modulation (16QAM), without any measurable optical-signal-to-noise-ratio (OSNR) penalty compared to a commercial PBS.

## 1. Introduction

Polarization manipulation is of great importance for integrated optical systems, in particular when it comes to interfacing rotationally symmetric optical fibers with degenerate polarization states to highly polarization-sensitive on-chip waveguides. In conventional optical systems, polarization manipulation usually relies on discrete optical elements such as polarization beam splitters (PBS) or waveplates made from birefringent materials. These devices offer high polarization extinction ratios and low insertion loss over a large spectral range. When used in integrated photonic systems, however, the viability of discrete polarization-manipulating elements is limited, e.g., by the required footprint and by the need for high-precision alignment of these elements with respect to on-chip optical circuits[8]. Alternatively, polarization-manipulating functionalities can be integrated into waveguide-based planar photonic circuits, exploiting, e.g., mode-selective directional or multimode interference couplers[9–12], polarization mode converters[13,14], waveguide gratings[15], waveguide structures with multi-layer cores[16,17], or more complicated structures obtained by inverse design techniques[18]. These devices can be efficiently realized in large quantities, but often require special fabrication steps[16,17] and are subject to limitations of the device geometry, dictated by conventional layer-by-layer microstructuring through 2D lithography and dry etching. Moreover, polarization manipulation in on-chip structures often relies on efficient coupling of light to the associated waveguides in the first place. In silicon photonics, grating-based polarization beam splitters can be directly integrated into the fiber-chip interface[19]. However, these structures are subject to limited bandwidth and still rely on high-precision active alignment of the single-mode fiber with respect to the on-chip grating coupler. More recently, 3D-printing based on multi-photon lithography has been exploited to realize PBS structures on the facets of single-mode fibers, exploiting polarization-sensitive diffraction gratings[20] and directional couplers within photonic-bandgap waveguides[21]. While these demonstrations already show the potential of 3D-printing for fabrication of PBS structures, the split signals are simply emitted into free space – without polarization rotation or coupling to further waveguide-based devices. In addition, the underlying physical effects employed in these structures fundamentally limit their bandwidth.

In this paper, we demonstrate that ultra-broadband 3D-printed waveguide-based polarization beam splitters and rotators open an attractive path towards polarization-manipulation in integrated optics. In our structures, polarization splitting is accomplished through adiabatic Y-branches of geometrically birefringent polymer waveguides with high-aspect-ratio cross sections and complemented by polarization rotation in waveguides that are twisted along the propagation direction. The structures can be directly incorporated into freeform chip-chip and fiber-chip connections[4,5], so-called photonic wire bonds. In our proof-of-concept experiments, we show monolithic structures that are 3D-printed on facets of single-mode fibers, comprising ultra-broadband polarization beam splitters, polarization rotators, and mode-field adapters. Our prototypes feature more than 11 dB polarization extinction ratios in a wavelength range between 1270 nm and 1620 nm, with vast potential for further improvement. To demonstrate the practical viability of the structures, we use them in a dual-polarization data-transmission experiment, in which we receive a dual-polarization 16-state quadrature amplitude modulation (16QAM) data stream at a symbol rate of 80 GBd and an aggregate data rate of 640 Gbit/s. We find that our 3D-printed PBS do not introduce any measurable optical-signal-to-noise-ratio (OSNR) penalty when compared to a commercially available fiber-coupled PBS. We believe that 3D-printed optical structures for polarization manipulation can replace costly assemblies of discrete micro-optical elements, thereby paving the path towards optical systems with unprecedented compactness and scalability.

## 2. PBS concept and simulations

The basic concept of 3D-printed polarization beam splitters (PBS) and polarization rotators (PR) in integrated optical assemblies is illustrated in Fig. 1. The device connects a rotationally symmetric single-mode fiber (SMF) with degenerate polarization states to a highly polarization-sensitive photonic integrated circuit (PIC). The illustrated assembly acts as a dual-polarization receiver for coherent communications, in which data signals in orthogonal polarization states of the SMF are split and independently detected using a pair of coherent optical receivers (Coh. Rx), which are fed by a joint local oscillator (LO). The PBS/PR can be merged with additional 3D freeform waveguide elements such as mode-field adapters into a single monolithic structure. This structure can be fabricated in a single exposure step by high-resolution 3D-laser lithography that exploits multi-photon polymerization in the focus of a pulsed femtosecond laser beam[2]. This offers the freedom to adapt the geometry of the 3D-printed structure to the positions of the adjacent optical device facets, thereby overcoming the need for high-precision mechanical alignment of the fiber with respect to the chip[4,5]. Note that the assembly illustrated in Fig. 1 represents only one example how 3D-printed polarization-manipulating elements can be used in integrated optics. In general, the structures can be printed on a wide range of optical devices, covering applications from optical communications and signal processing[22,23] to optical metrology,[24] imaging,[25] and quantum optics[26].

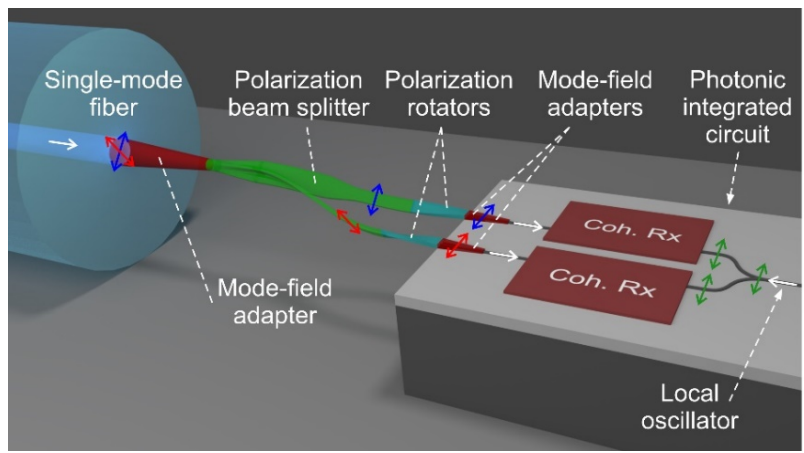

**Figure 1: Concept of a 3D-printed polarization beam splitter and rotator in an integrated optical assembly (not drawn to scale).** The device connects a rotationally symmetric single-mode fiber (SMF) with degenerate polarization states (red and blue arrows) to a photonic integrated circuit (PIC) with highly polarization-sensitive waveguides. As an example of high practical interest, we illustrate a dual-polarization receiver for coherent communications, in which data signals in orthogonal polarization states are split and independently detected using a pair of coherent optical receivers (Coh. Rx) which are fed by a joint local oscillator (LO). The polarization beam splitter (PBS) and the polarization rotators (PR) can be merged with additional 3D freeform waveguide elements such as mode-field adapters to form a single monolithic structure. This structure can be fabricated in a single exposure step by high-resolution 3D-laser lithography, thereby offering the freedom to adapt the geometry of the 3D-printed structure to the positions of the various optical device facets.

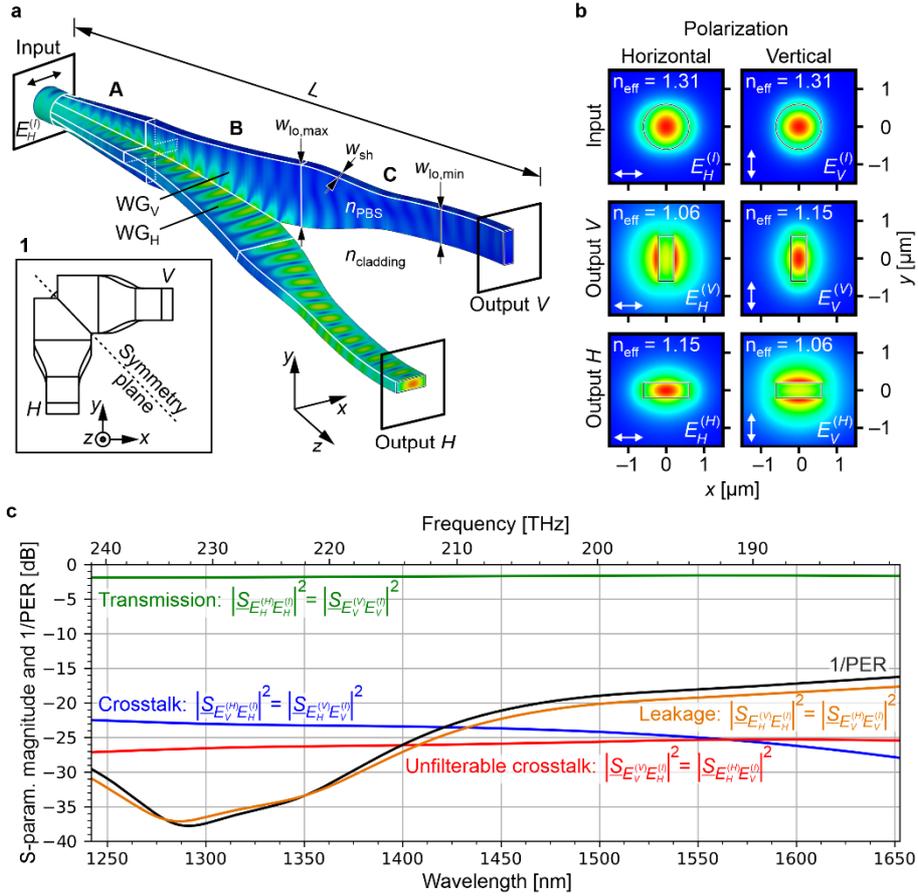

**Figure 2: Concept and design of 3D-printed waveguide-based PBS. a**, 3D model of the PBS, comprising an input waveguide port with a circular cross section and a pair of output waveguide ports with rectangular cross sections of high aspect ratio. The two orthogonally polarized modes at the input port are denoted as $E_H^{(I)}$ and $E_V^{(I)}$, whereas $E_H^{(V)}$ refers to the horizontally and $E_V^{(V)}$ to the vertically polarized mode at the vertical output $V$, while $E_V^{(H)}$ denotes the vertically and $E_H^{(H)}$ the horizontally polarized mode at the horizontal output $H$. The PBS consists of three segments denoted by A, B, and C. Within Segment A, the circular cross section at the input port is adiabatically morphed into a cross-shaped cross section. Within Segment B, the structure can be represented by two spatially overlapping partial waveguides $WG_H$ and $WG_V$ with high-aspect-ratio rectangular cross sections, which are gradually separated to drag the strongly guided eigenmodes into the two distinct waveguides at the input of Segment C. The 3D rendering of the structure also depicts the simulated electric field distribution for a horizontally polarized excitation $E_H^{(I)}$ at the input port. The PBS exhibits full geometrical symmetry with respect to a plane that is oriented at 45° between the horizontal and the vertical direction, see Inset 1. The refractive index of the 3D-printed PBS core region amounts to $n_{PBS} = 1.53$, and the cladding material is air, $n_{cladding} = 1$. **b**, Electric field plots ($|\mathbf{E}|$) of the fundamental modes for both polarizations at all three ports of the PBS. The arrows indicate the orientation of the dominant transverse component of the electric field. The strongly guided target modes $E_H^{(H)}$ and $E_V^{(V)}$ at the horizontal and vertical output exhibit a higher effective index and a stronger confinement to the rectangular core than the undesired modes $E_V^{(H)}$ and $E_H^{(V)}$. **c**, Simulated wavelength dependence of the squared magnitudes of complex scattering parameters (S-parameters) and the reciprocal of the polarization extinction ratio (1/PER) of the PBS on a logarithmic scale. The transmission is better than −2.0 dB with a maximum of approximately −1.6 dB near $\lambda = 1550$ nm. The reciprocal of the polarization extinction ratio (1/PER), and the spurious coupling $\left|\underline{S}_{E_V^{(H)}E_H^{(I)}}\right|^2 = \left|\underline{S}_{E_H^{(V)}E_V^{(I)}}\right|^2$, $\left|\underline{S}_{E_H^{(H)}E_V^{(I)}}\right|^2 = \left|\underline{S}_{E_V^{(V)}E_H^{(I)}}\right|^2$, and $\left|\underline{S}_{E_H^{(V)}E_H^{(I)}}\right|^2 = \left|\underline{S}_{E_V^{(H)}E_V^{(I)}}\right|^2$ between input and output modes are below −16 dB over the 400 nm wide wavelength range. These parameters can be further reduced for smaller wavelength ranges. Details on extracting the PER from the simulations can be found in Supplementary Information Section S2.

The working principle of our 3D freeform waveguide-based PBS is illustrated in Fig. 2. Figure 2a depicts a 3D rendering of the three-port device, comprising an input waveguide port with a circular cross section and a pair of output waveguide ports with rectangular cross sections of high aspect ratio. In the following, the input port is denoted by a superscript $(I)$, whereas superscripts $(H)$ and $(V)$ refer to the output ports with horizontally and vertically oriented rectangular cross section, see Fig. 2a. Note that the device is fully bidirectional and can also be used as a polarization beam combiner, where the two rectangular waveguide ports $H$ and $V$ are used as inputs, whereas the circular port $I$ acts as output.

The PBS consists of three segments, denoted by A, B, and C, where Segment A is directly adjacent to the input port. Due to its circular cross section, the input port has two degenerate fundamental modes of orthogonal polarizations with identical effective refractive indices $n_{\text{eff}}$. Without loss of generality, we select the two basis modes at the input with the dominant transverse component of the electric field aligned in the vertical and horizontal direction as defined by the two output ports, and we denote them as $E_H^{(I)}$ and $E_V^{(I)}$ respectively, see first row of Fig. 2b for the associated mode-field distributions. Within Segment A, the circular cross section at the input port is adiabatically morphed into a cross-shaped cross section at the transition to Segment B. At the transition between Segment A and Segment B the waveguide can be represented by two spatially overlapping partial waveguides $WG_H$ and $WG_V$ with high-aspect-ratio rectangular cross sections. Due to the adiabatic transition from a circular cross section to a cross-shaped one in Segment A, the two degenerate basis modes denoted as $E_H^{(I)}$ and $E_V^{(I)}$ are transformed into the strongly guided modes of these rectangular partial waveguides. In this context, the term "strongly guided" denotes a mode of a rectangular waveguide which is polarized along the long side of the rectangle. This mode exhibits a higher effective refractive index than its weakly guided counterpart that is polarized along the short side of the rectangular waveguide core. In Segment B, the partial waveguides $WG_H$ and $WG_V$ are gradually separated, thereby dragging the corresponding strongly guided eigenmodes into the two distinct waveguides at the input of Segment C.

Within Segment C, the two distinct output waveguides can be adiabatically tapered and bent to suppress unwanted higher-order modes and to route the waveguides to the two output ports $V$ and $H$. Further 3D-printed freeform waveguide structures can be directly connected to these output ports, e.g., for polarization rotation, see Fig. 1. The second and the third row of Fig. 2b show the various mode-field profiles at the output, where $E_H^{(V)}$ refers to the horizontally and $E_V^{(V)}$ to the vertically polarized mode at the vertical output $V$, whereas $E_V^{(H)}$ denotes the vertically and $E_H^{(H)}$ the horizontally polarized mode at horizontal output $H$. In an ideal device, the power of the degenerate $E_H^{(I)}$ and $E_V^{(I)}$ is completely coupled to the strongly guided modes $E_H^{(H)}$ and $E_V^{(V)}$, whereas the weakly guided modes $E_H^{(V)}$ and $E_V^{(H)}$ are not excited.

To estimate the performance of the proposed PBS, we perform numerical simulations of the full 3D structure, see Methods for details. For good performance, the aspect ratio of the rectangular waveguide cross sections should be as high as possible while staying compatible with the resolution of the 3D printing system. In the simulated structure, the short side of the rectangle was chosen to be $w_{\text{sh}} = 400$ nm, whereas the long side varied in the range $w_{\text{lo}} = (1.2 \ldots 2.2)\,\mu\text{m}$. The length of the structure including all three segments amounts to $L = 21\,\mu\text{m}$. The PBS features full geometrical symmetry, see Inset 1 of Fig 2a, which results in symmetrical relationships between the port modes. In Fig. 2a, we illustrate the magnitude of the E-field at a wavelength of 1550 nm for horizontal polarization at the input – the E-field distribution for vertical input polarization is obtained by reflection about the symmetry plane. To describe the coupling between the various modes at the input and the output ports, we use complex-valued scattering parameters (S-parameters) $\underline{S}_{IJ}$. In this description, $\underline{S}_{IJ}$ refers to the amplitude transmission from $J$ to $I$ where $I, J \in \left\{ E_H^{(I)}, E_V^{(I)}, E_H^{(H)}, E_V^{(H)}, E_H^{(V)}, E_V^{(V)} \right\}$ denote the various modes at the ports of the device. The results for the various simulated power coupling coefficients, which are obtained by squaring the magnitudes of the corresponding S-parameters are shown in Fig 2c. Evaluating the power transmission $\left| \underline{S}_{E_H^{(H)} E_H^{(I)}} \right|^2 = \left| \underline{S}_{E_V^{(V)} E_V^{(I)}} \right|^2$ from the input to the desired mode of the respective output port, we find an insertion loss of less than 2 dB over the entire wavelength range between 1250 nm and 1650 nm, with a minimum of 1.6 dB near 1550 nm, see green line in Fig. 2c. For each polarization at the input port, we further extract the power that is coupled to the undesired mode at the corresponding "correct" output port, which is quantified by the crosstalk $\left| \underline{S}_{E_V^{(H)} E_H^{(I)}} \right|^2 = \left| \underline{S}_{E_H^{(V)} E_V^{(I)}} \right|^2$, and which is below –22 dB throughout the simulated wavelength range, see blue line in Fig. 2c. Note that this crosstalk can be suppressed by subsequent polarization filtering. We further calculate the power that is coupled to the desired output modes $E_H^{(H)}$ and $E_V^{(V)}$, from the "wrong" input modes $E_V^{(I)}$ and $E_H^{(I)}$, respectively, and that cannot be suppressed by subsequent polarization filters. This unfilterable crosstalk $\left| \underline{S}_{E_H^{(H)} E_V^{(I)}} \right|^2 =$ $\left|\underline{S}_{E_V^{(V)}E_H^{(I)}}\right|^2$ is represented by the red line in Fig. 2c and is below –25 dB throughout the 400 nm-wide wavelength range of interest. We also extract the polarization leakage $\left|\underline{S}_{E_H^{(V)}E_H^{(I)}}\right|^2 = \left|\underline{S}_{E_V^{(H)}E_V^{(I)}}\right|^2$, which, for a given polarization at the input port quantifies the power coupled to undesired polarization at the "wrong" output port, thereby maintaining its polarization direction. For our structure, the polarization leakage is below –17 dB throughout the simulated wavelength range, see orange line in Fig. 2c, and can be further suppressed by subsequent polarization filters. Finally, we extract the polarization extinction ratio (PER), i.e., the ratio of the maximum and the minimum power observed in both modes of an output port when varying the excitation at the input over all possible polarization states. For each of the output ports, the PER can be obtained from a singular-value decomposition of the corresponding Jones matrix, see Supplementary Information Section S2 for details. We find that the PER is better than 16 dB within the investigated wavelength range and shows a strong increase towards longer wavelengths. Note that the PER and polarization leakage are better than 30 dB over the wavelength range between 1250 nm and 1365 nm, and that this wavelength range of high performance can be shifted by adapting the design of the structure.

## 3. Experiments

To experimentally prove the viability of our concept, we fabricate a series of 3D-printed PBS that are directly connected to the cores of single-mode fibers (SMF). We characterize the performance of these devices and finally use them as part of a receiver in a polarization-division multiplexing (PDM) data transmission experiment.

In a first experiment, we fabricate a series of free-standing PBS on the facets of an SMF array and measure the performance through an infra-red-sensitive microscope (IR microscope), see Fig. 3a. To ensure low-loss coupling to the SMF core, the structures are equipped with adiabatic mode-field adapters that are attached to Segment A of the PBS. A scanning-electron microscope (SEM) image of the PBS and the mode-field adapter are shown in Fig. 3b. Light is fed to the SMF by a laser emitting at a wavelength of $\lambda = 1510$ nm, and subsequent polarization controller. At the two PBS outputs, the light is radiated into free space and picked up by the IR microscope. The centers of the white circles in the images of Fig. 3c match the centers of corresponding PBS output ports, and the areas of the circles denote the areas that have been considered in calculating the corresponding power levels. In a first set of measurements, we show that radiated light can be switched between the two output ports of the PBS by varying the polarization at the input, see Column 1 of Fig. 3c. Specifically, Subfigures 1.1 and 2.1 refer to the cases where the polarization controller was adjusted for maximum radiation from output port $V$ and $H$, having vertically and horizontally oriented waveguide cross sections, respectively. In both cases, we measure the ratio $\Gamma$ of the optical power at the targeted output port to the residual power emitted at the respective other port, which amounts to 9.8 dB and 9.7 dB, respectively. We also adjust the input polarization to yield equal power at both ports, see Subfigure 3.1 in Fig. 3c. To check the polarization states of the light radiated from the two outputs, we repeat the experiment with a vertically and horizontally oriented polarization filter (PF) between the microscope objective and the IR camera, see Columns 2 and 3 of Fig. 3c. The PF has an extinction ratio of more than 34 dB. Assuming an excitation with pure vertical polarization in Row 1 of Fig. 3c, the vertically oriented PF in Subfigure 1.2 suppresses the spurious horizontal polarization at Port $V$, which corresponds to the crosstalk $\underline{S}_{E_H^{(V)}E_V^{(I)}}$, as well the spurious horizontal polarization at Port $H$, which represents the unfilterable crosstalk $\underline{S}_{E_H^{(H)}E_V^{(I)}}$. The measured power ratio $\Gamma$ of the emitted light after the PF amounts to 12.1 dB and corresponds to the ratio $\left|\underline{S}_{E_V^{(V)}E_V^{(I)}}\right|^2 \Big/ \left|\underline{S}_{E_V^{(H)}E_V^{(I)}}\right|^2$ of the power transmission at Port $V$ and the leakage at Port $H$. The measured ratio is smaller than the approximately 18 dB that would be expected from the simulation results, see Fig. 2c. We attribute the deviations to geometrical inaccuracies of the fabricated structure. In Subfigure 2.2, the polarization controller is adjusted for maximum radiation from output *H*, but the PF is oriented vertically, such that only spurious horizontal polarizations at both outputs, $\left|\underline{S}_{E_H^{(H)}E_V^{(I)}}\right|^2$ and $\left|\underline{S}_{E_H^{(V)}E_V^{(I)}}\right|^2$ can be seen on the IR camera. As expected, the camera image does not show any significant power. In Subfigure 3.2, where light exits both output arms of the PBS, the light radiated from Port *H* is completely suppressed by the vertically oriented PF, whereas the light radiated from Port *V* does not experience a significant attenuation. The same experiment is repeated with a horizontally oriented PF, see Column 3 of Fig. 3c, thereby essentially reproducing the findings described for the data in the second column. This simple experiment demonstrates that the device qualitatively works as expected.

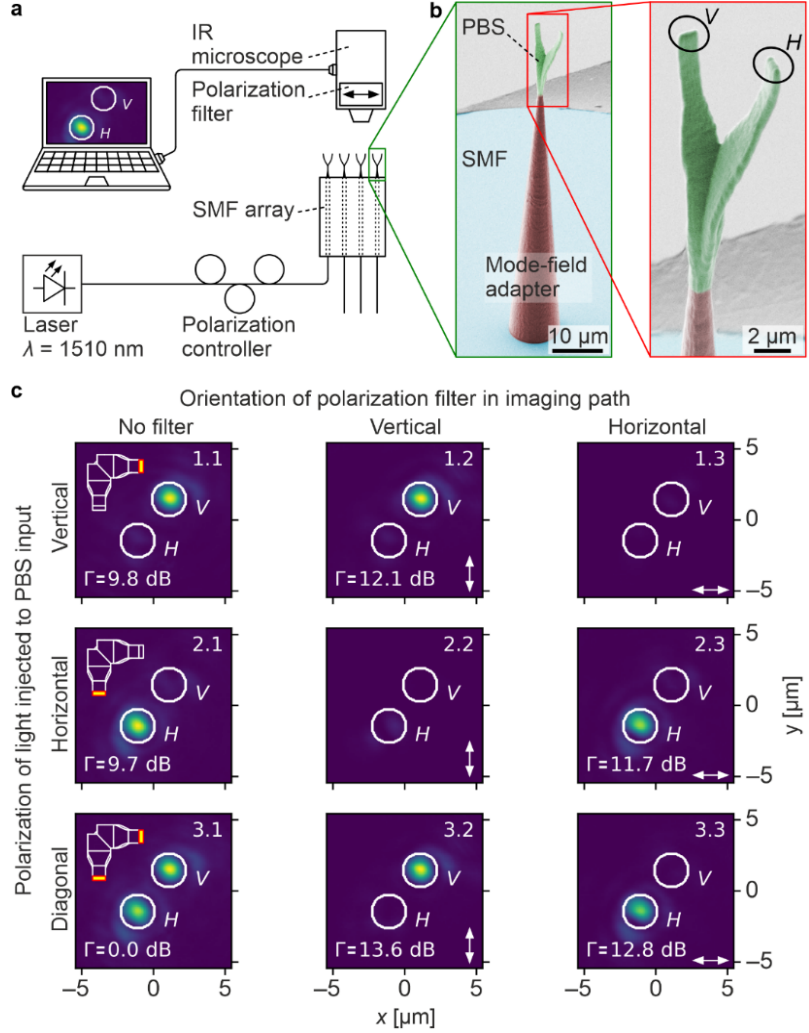

**Figure 3: Characterization of 3D-printed PBS using an infra-red-sensitive (IR) microscope. a,** Experimental setup: As test structures, we use a series of PBS that are 3D-printed on the facets of a single-mode fiber (SMF) array. Light at a wavelength of 1510 nm is fed to the devices by a laser and a subsequent polarization controller. Light emitted from the PBS is characterized by an IR microscope equipped with polarization filter (PF). **b,** Scanning-electron microscopy (SEM) images of a fabricated structure on the fiber array. A linear taper structure, shaded in red, is used at the input of the PBS to adapt the mode-field diameter of the SMF to the one of the PBS input. Within the PBS, which is illustrated in green, the light is split into two orthogonal polarizations and emitted from the outputs (*V* and *H*) towards the IR microscope. Colors were added by image processing. **c,** Recordings on the IR microscope for different combinations of input polarization states, indicated by the different rows: Row 1 – vertical input polarization only, Row 2 – horizontal input polarization only, and Row 3 – both vertical and horizontal input polarizations. The columns correspond to the measurement of the radiated power without (Column 1) and with vertically and horizontally oriented polarization filter (Columns 2 and 3, respectively) in the imaging path of the IR microscope. The output power of each port is estimated by integrating the measured intensity over the areas within the white circles, and a power ratio $\Gamma$ in dB is calculated by dividing the larger by the smaller power. A top view of the PBS structure and the respective "active" output port for each row is additionally illustrated in Column 1. The orientation of the polarization axis of the PF is illustrated by the double arrows in the lower right-hand corner of the displays in Columns 2 and 3.

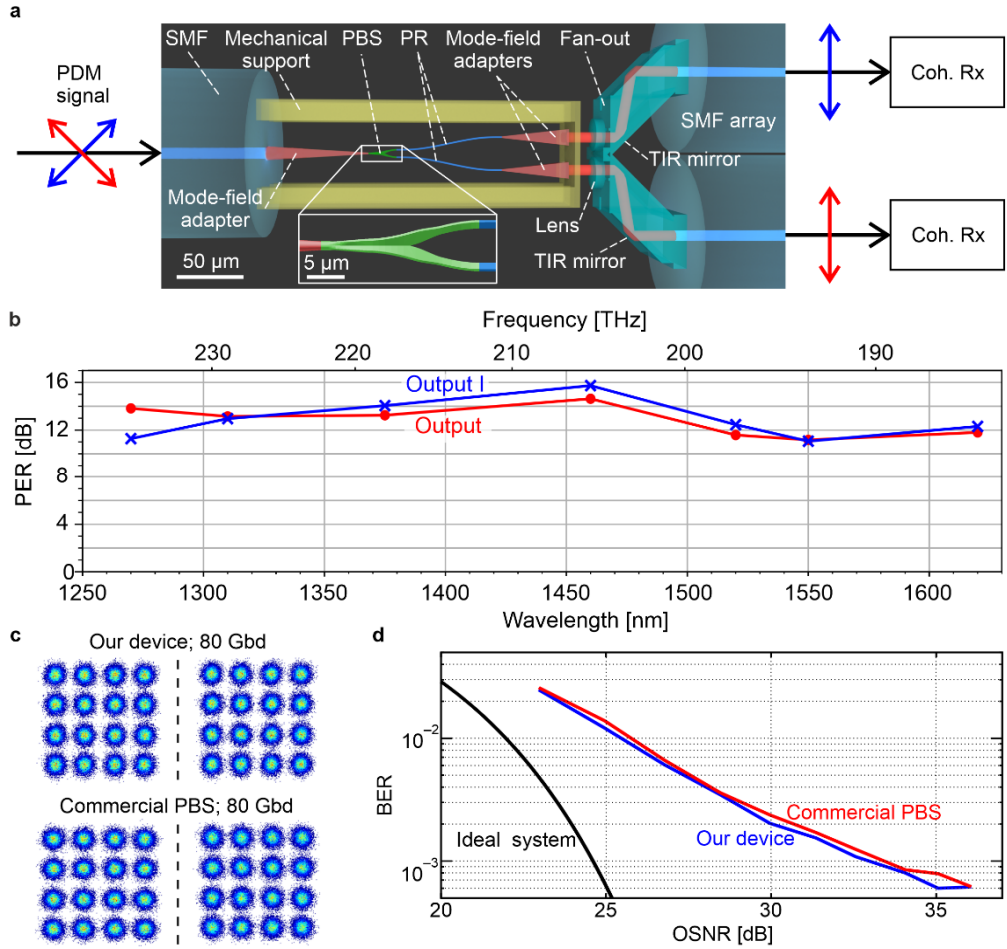

**Figure 4: Experimental setup and results of proof-of-concept data transmission experiment. a,** Simplified experimental setup: The polarization-division-multiplexed (PDM) 16QAM signal is fed to an SMF having a mode-field adapter and a 3D-printed polarization beam splitter (PBS) on its facet. The PBS is additionally equipped with 3D-printed polarization rotators (PR) in the form of twisted waveguides, which rotate the polarizations in both output ports to an identical direction. We simultaneously probe the two output signals by a fan-out structure that is 3D-printed on a second SMF array. The fan-out consists of two lenses and two pairs of total-internal-reflection (TIR) mirrors to adapt the 25 μm pitch of the PBS/PR outputs to the 127 μm pitch of the SMF in the array. The signals are subsequently decoded by a pair of commercial coherent receivers (Coh. Rx). To benchmark our device, we repeat the experiment by replacing the PBS/PR assembly and the fan-out by a commercial fiber-coupled PBS. **b,** Measurement of the PER for both outputs: The PER is better than 11 dB in the wavelength range (1270 … 1620) nm, which was only limited by the tuning range of the underlying laser sources. **c,** Constellation diagrams of received 80 GBd 16QAM signals for an optical signal-to-noise ratio (OSNR) of 36 dB. Upper row: experiment with our device. Lower row: experiment with the commercial PBS. **d,** Bit-error-ratio (BER) vs. OSNR. **Black:** Theoretical curve for an ideal transmission system. **Blue:** Experiment with our 3D-printed PBS/PR assembly. **Red:** Experiment with the commercial PBS. Our device does not introduce an OSNR penalty with respect to the commercial PBS. At a BER of $1.25 \times 10^{-2}$, which corresponds to the threshold of forward error correction with 15 % coding overhead, our transmission setup exhibits an implementation penalty of approximately 3 dB, see Supplementary Information Section S4 for details.

In a second experiment, we further test our PBS structures by measuring the PER over a broad range of wavelengths. To this end, the polarization at the input of the structure is varied randomly by a polarization scrambler, while the Stokes vector and the power at the device output are continuously recorded by a polarimeter, see Supplementary Information Sections S1–S3 for details. The measurement is repeated for each of the output ports, thereby revealing the output polarization state of maximum and minimum power transmission as well as the associated PER. The PBS test structure used in this experiment is again 3D-printed on the facet of an SMF array, which is connected to the polarization scrambler. At the output, the structure is equipped with a pair of polarization rotators (PR), realized by rectangular waveguides that are twisted by 45° along the propagation direction[27,28], thus providing identical polarizations at both ports, see Fig. 1 and Fig. 4a. For better probing of the output, the structure is equipped with adiabatic mode-field adapters that are held by a table-like mechanical support structure, Fig. 4a. The output ports can hence be individually read out by an SMF, see Supplementary Information Section S1 for details of the experiment. We find a measured PER better than 11 dB in the wavelength range (1270 … 1620) nm, which was only limited by the tuning range of the underlying laser sources, see Fig. 4b. The measured insertion losses at 1550 nm for Output I and Output II correspond to 4.4 dB and 3.8 dB, respectively, including the loss of the PBS, of the subsequent PR, and of the adiabatic mode-field adapters at the input and the output of the device, see Supplementary Information Section S1 for details.

To demonstrate the technical viability of 3D-printed PBS, we finally perform a polarization division multiplexing (PDM) data-transmission experiment that emulates the application scenario illustrated in Fig. 1. The experimental setup and the results are shown in Fig. 4a, c, and d. Details of the experimental setup can be found in the Supplementary Information Section S4. The experiment relies on a PBS/PR combination as used in the previous experiment, complemented by an additional fan-out structure, see Fig. 4a. This fan-out structure is 3D-printed on a second fiber array and allows to simultaneously couple both PBS outputs to a pair of outgoing SMF with a standard pitch of 127 μm. The fan-out structure is equipped with two lenses with a pitch of 25 μm that pick-up light from the two PBS outputs, followed by a pair of total-internal-reflection (TIR) mirrors in each path to adjust the lateral offset of the beams at the output. At $\lambda = 1550$ nm, the measured insertion losses of the two channels of the fan-out are 1.2 dB and 1.9 dB, with a crosstalk between channels below –46 dB.

In the transmission experiment, we launch a 16QAM PDM signal at a symbol rate of 80 GBd and a center wavelength of $\lambda = 1550$ nm to the PBS, and we feed the two output signals of the PBS/PR assembly to a pair of coherent receivers, see Supplementary Information Section S3 for details. We perform the experiment both with our 3D-printed PBS assembly and with a commercially available fiber-coupled PBS having a PER in excess of 30 dB. In both cases, we sweep the optical signal-to-noise ratio (OSNR) at the input of the PBS and record the constellation diagrams along with corresponding bit error ratios (BER), see Figs. 4c and 4d. At a BER of $1.25 \times 10^{-2}$, our transmission setup exhibits an OSNR penalty of approximately 3 dB with respect to an ideal transmission system, Fig. 4d, see Supplementary Information Section S4 for details. We find that the 3D-printed PBS/PR assembly does not introduce any additional OSNR penalty with respect to the commercial PBS, although the PER differ vastly. This may be attributed to the fact that the polarization-sensitive mixing of the data signals with the local oscillator of the coherent optical receiver in combination with the digital polarization demultiplexing algorithms of the PDM receiver can easily compensate for the finite PER of our 3D-printed PBS. Hence, even though there is still room for improving the performance of our 3D-printed PBS/PR assemblies, the current devices already offer an attractive route towards highly scalable ultra-compact dual-polarization receivers as shown in Fig. 1.

## 4. Summary

We demonstrated 3D-printed waveguide-based polarization beam splitters (PBS) that can be efficiently integrated into chip-chip and fiber-chip interfaces of optical assemblies. The devices rely on adiabatic Y-branches of geometrically birefringent waveguides with high-aspect-ratio cross sections and can be complemented by polarization rotators (PR) that exploit twisted 3D freeform waveguides. The PBS/PR can be efficiently fabricated by direct-write two-photon lithography together with other 3D-printed elements such as photonic wire bonds[4,5], 3D-printed microlenses[7], or 3D-printed waveguide overpasses[29]. In our proof-of-concept experiments, we demonstrate broadband operation in the wavelength range of (1270 … 1620) nm, limited only by the available equipment. We further prove the practical viability of the concept in a high-speed data transmission experiment, where our 3D-printed PBS/PR assemblies are used for separating polarization-multiplexed data signals at the receiver. We find that the quality of the received signals is on par with that obtained by a conventional high-performance fiber-coupled PBS. While the concept leaves room for further optimization, we believe that 3D-printed PBS/PR can pave the path towards efficient polarization manipulation in integrated optical systems with unprecedented compactness and scalability.

## Methods

**Simulations**: For the simulations shown in Fig. 2, the 3D structure of the PBS was modeled using a commercially available 3D numerical time-domain solver (CST Studio Suite, Dassault Systèmes, Vélizy-Villacoublay, France). The final design of the PBS was the result of a manual optimization based on several parameter sweeps.

**Fabrication**: All 3D-printed structures were fabricated using a home-built two-photon lithography system equipped with a 63× microscope objective lens (numerical aperture 1.4, field number 25 mm) and galvanometer mirrors for rapid beam movement in the lateral directions. As a lithography light source, we use a fs-laser with a pulse length of less than 80 fs (CFiber 780 Femtosecond Fiber Laser, Menlo Systems GmbH, Planegg, Germany) and a repetition rate of 100 MHz. The lithography system is equipped with a dedicated control software that allows for precise localization of the optical fiber core as well as for automated fabrication of the PBS with high shape fidelity. The system is equipped with a confocal imaging unit using the lithography laser and its beam deflectors for the acquisition of 3D images that are perfectly aligned to the lithography coordinate system and hence to any lithographically fabricated structures. For confocal imaging, the laser power is reduced to avoid any unwanted polymerization in the photoresist. In the lithography process, the liquid negative-tone photoresist (Nanoscribe IP-Dip, refractive index $n = 1.52$ at 780 nm, unexposed; see also Ref. [30]) simultaneously acts as an immersion medium for the objective lens. Unexposed photoresist is removed in a two-step development process using propylene-glycol-methyl-ether-acetate (PGMEA) as a developer for 20 min, followed by rinsing in isopropyl alcohol (2-propanol).

**Trajectory planning and fiber-to-PBS interface:** For the polarization rotators and the output waveguides, careful planning of the 3D trajectory is important to ensure efficient coupling between the PBS and other optical structures. To this end, we use a parametrized trajectory and optimize it for low curvature and hence low radiation loss. Along this trajectory, the waveguide cross section is extruded to form a 3D model of the structure that is then lithographically fabricated. Low-loss coupling between PBS and the standard single-mode fiber (Corning SMF-28) at its input is achieved by a linearly tapered mode-field adapter, designed for a mode-field diameter (MFD) of $(10.3 \pm 0.4)$ µm at 1550 nm at the fiber side. The MFD is defined as the diameter at which the intensity has dropped to $1/e^2$ of its maximum value measured in the center of the fiber core. The methods are derived from the photonic wire bonding process, details on which can be found in Ref. [5].

**Characterization using an IR microscope:** For characterization of the 3D-printed PBS in Fig. 3, we use an IR camera (Goldeye G-032 SWIR, Allied Vision) attached to a microscope (DMRXA with a variable zoom unit DMRD, Leica/Leitz) that is equipped with an IR objective (LMPlan IR 100×/0.80NA, Olympus). An optional rotatable linear polarizer (LPIREA100-C, Thorlabs, PER > 34 dB at 1550 nm) can be inserted into the infinity-optical beam path of the microscope. Laser light generated by a tunable external-cavity laser (IQS-2600B, EXFO) is injected into the SMF, and the polarization is adjusted by a standard fiber-based polarization controller. Each acquired image is corrected for the background signal that is seen with the laser turned off.

**Polarization extinction ratio (PER) measurement:** The PER is measured by using an optical component analyzer (Keysight N7788B), which comprises an integrated fast polarization controller and a polarimeter. The polarization controller randomly scrambles the state of polarization, thereby uniformly covering the whole Poincaré sphere. The polarization state and the power at the output of the PBS structure are measured simultaneously by the polarimeter, see Supplementary Information Section S1 for details. The PER can be extracted from these measurements, see Supplementary Information Section S3 for details. The PER is measured at seven discrete wavelengths between 1270 nm and 1620 nm, using three different tunable laser sources (Ando AQ4321D, TUNICS T1005-HP, Agilent 81600B).

**Data transmission experiment:** In our data transmission experiments, we used four output channels from an AWG (Keysight M8196A) to generate the drive signals for the dual-polarization IQ modulator, see Supplementary Information Section S4 for a sketch of the underlying experimental setup. The signals are derived from random bit patterns with different seeds, such that each polarization carries uncorrelated data, and are pre-distorted to compensate for the measured frequency response of the transmitter. For the optical-signal-to-noise-ratio (OSNR) sweep, band-limited amplified spontaneous-emission (ASE) noise is generated by an ASE source (Orion Laser Technology ASE-C/CL) and added to the data signal. The noise-loaded signal is then fed to the receiver, which comprises an erbium-doped fiber amplifier (EDFA) for pre-amplification, followed by a bandpass filter (full width at half maximum 1 nm) to suppress out-of-band amplified spontaneous emission (ASE) noise. The signal is then fed to the 3D-printed PBS/PR shown in Fig. 4a, where the two orthogonal polarization states are split and rotated. The two partial signals are then detected using a coherent heterodyne scheme, where the optical local oscillator tone (LO, Keysight N7714A) is tuned to the edge of the signal spectrum and where two balanced photodiodes (Finisar BPDV2150RQ) remove both signal-signal and LO-LO interference, see Supplementary Information Section S4 for details. The outputs of the photodiodes are digitized by a 256 GSa/s real-time oscilloscope (Keysight UXR1004A) and recorded for offline digital signal processing (DSP). In a first DSP step, the signals are shifted in frequency by the difference between the carrier and the LO tone. After timing recovery, a $2 \times 2$ MIMO equalizer is used for polarization de-multiplexing, followed by carrier recovery, see Supplementary Information Section S4 for details. Finally, the signals go through a least-mean-square equalizer before being finally decoded.

## Acknowledgments

This work was supported by the Deutsche Forschungsgemeinschaft (DFG, German Research Foundation) in the framework of the Collaborative Research Center (CRC) Wave Phenomena (SFB 1173, project-ID 258734477) and under Germany's Excellence Strategy via the Excellence Cluster 3D Matter Made to Order (EXC-2082/1 – 390761711), by the Bundesministerium für Bildung und Forschung (BMBF) Project PRIMA (# 13N14630), by the European Research Council (ERC Consolidator Grant 'TeraSHAPE', # 773248), by the H2020 project TeraSlice (# 863322) and the Photonic Packaging Pilot Line PIXAPP (# 731954), by the Alfried Krupp von Bohlen und Halbach Foundation, by the Karlsruhe School of Optics and Photonics (KSOP), and by the Karlsruhe Nano-Micro Facility (KNMF). A.N. was supported by the Erasmus Mundus joint doctorate programme Europhotonics (grant number 159224-1-2009-1-FR-ERA MUNDUS-EMJD).

*This page intentionally left blank.*

# Ultra-broadband polarization beam splitter and rotator based on 3D-printed waveguides

## Supplementary Information

A. Nesic[1†*], M. Blaicher[1,2†], P. Marin-Palomo[1], C. Füllner[1],
S. Randel[1], W. Freude[1], C. Koos[1,2,3*]

*[1]Institute of Photonics and Quantum Electronics (IPQ), Karlsruhe Institute of Technology (KIT), Karlsruhe, Germany*
*[2]Institute of Microstructure Technology (IMT), Karlsruhe Institute of Technology (KIT), Eggenstein-Leopoldshafen, Germany*
*[3]Vanguard Automation GmbH, Karlsruhe, Germany*
*†These authors contributed equally to this work.*
*\*e-mail: aleksandar.nesic@kit.edu, christian.koos@kit.edu*

## S1. Measurement of polarization extinction ratio (PER)

We test our PBS structures by measuring the PER over a broad range of wavelengths. To this end, the polarization at the input of the structure is varied randomly by a polarization scrambler, while the Stokes vector and the power at the device output are continuously recorded by a polarimeter, see Fig. S1a for a sketch of the associated setup. The measurement was performed with a commercially available optical component analyzer (Keysight N7788B) and was repeated for each of the output ports, thereby revealing the output polarization state of maximum and minimum transmission as well as the associated PER. The PBS test structure used in this experiment is 3D-printed on the facet of an SMF array which is connected to the polarization scrambler. At the output, the structure is equipped with a pair of polarization rotators (PR), realized by rectangular waveguides that are twisted by 45° along the propagation direction[1,2], thus providing identical polarizations at both ports, see Fig. S1b. For better probing of the output, the structure is equipped with adiabatic mode-field adapters that are held by a table-like mechanical support structure, see Fig. 4a of the main manuscript, and that can be individually probed by moving an SMF to the respective port. Note that, due to the unknown polarization rotation in the SMF, our measurement only allows to determine the exact polarization state at the input of the polarimeter, but not at the output ports of the PBS/PR. This needs to be considered when evaluating the measurement data, see Supplementary Section S3 for details. Note also that the two output ports of our structure are only separated by 25 μm, and we may hence assume that the polarization rotation in the SMF does not change significantly when moving the SMF between the ports. For an ideal device, the two ports should thus exhibit maximum transmission at identical output polarization states.

The measurement results obtained from our test structure at a wavelength of $\lambda = 1460$ nm are depicted in Fig. S1c. For this measurement, the input polarization state was scanned across 20 000 points uniformly distributed on the Poincaré sphere. The plot shows the measured Stokes states on the Poincaré sphere in Mollweide projection, colored by normalized transmitted power. For each of the two device outputs, we find a predominant polarization state, which we mark by $\mathbf{s}_{\mathrm{out,pass,1}}$ and $\mathbf{s}_{\mathrm{out,pass,2}}$ in Fig. S1c. These states correspond to the polarization that would be transmitted to the respective output of a perfect PBS. For a real device with finite PER, the output polarization states $\mathbf{s}_{\mathrm{out,pass,1}}$ and $\mathbf{s}_{\mathrm{out,pass,2}}$ exhibit the highest power transmission. At the same time, the measured output polarization states are concentrated around $\mathbf{s}_{\mathrm{out,pass,1}}$ and $\mathbf{s}_{\mathrm{out,pass,2}}$ in case the input polarization is randomly varied. Note that, for simplicity, we rotated all measured Stokes vectors such that $\mathbf{s}_{\mathrm{out,pass,1}}$ is oriented along the $s_1$-direction (latitude 0° and longitude 0°), which corresponds to a linear polarization in horizontal direction, while $\mathbf{s}_{\mathrm{out,pass,2}}$ is on the equator of the Poincaré sphere, corresponding to a linear polarization at a certain angle $\psi$ with respect to the horizontal direction. Note also that the transformation of the measured output polarizations to linear polarization states is somewhat arbitrary since the true polarization transformation in the output fiber is unknown. Still, we extract only a slight angle deviation of $\psi = -8.4°$ of the two equivalent linear polarization states, indicating a fairly good performance of the polarization rotators.

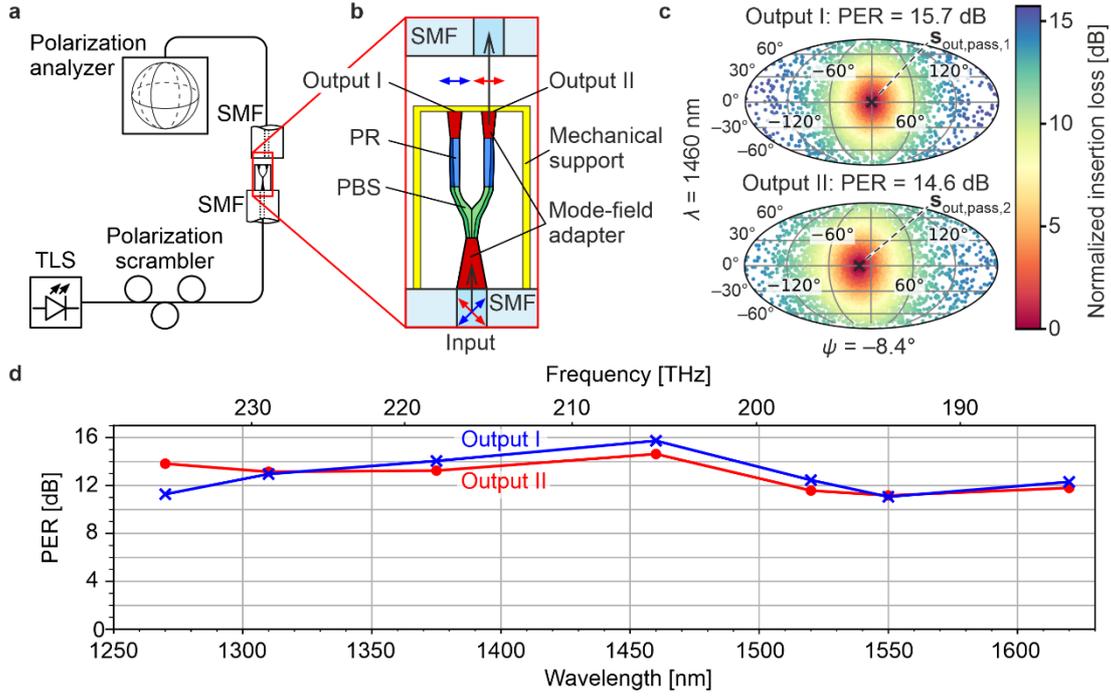

**Figure S1: Measurement of the polarization extinction ratio (PER) of the PBS with attached polarization rotators (PR). a,** Experimental setup: The PBS is 3D-printed together with the PR and additional mode-field adapters on the facet of an SMF, which is connected to a polarization scrambler. The two output ports are probed by a movable SMF, which is attached to a polarization analyzer. The polarization at the input is scrambled randomly, and the power and the Stokes vector of the output polarization state is measured at both outputs. **b,** Schematic rendering of the PBS (green) with attached PR (blue) and mode-field adapters (red), which are attached to a table-like mechanical support structure (yellow). Orthogonal polarization states (blue and red arrows) at the input port are separated to identical polarization states at the output of the structure. **c,** Measured output Stokes states on the Poincaré sphere in Mollweide projection, colored by normalized transmitted power. For simplicity, we rotate all measured Stokes vectors such that the polarization state with highest transmitted power at Output I, $\mathbf{s}_{\mathrm{out,pass,1}}$, is oriented along the $s_1$-direction of the Poincaré sphere (0° longitude and 0° latitude), which corresponds to a linear polarization in horizontal direction, while the predominant polarization state at Output II, $\mathbf{s}_{\mathrm{out,pass,2}}$, is on the equator of the Poincaré sphere, corresponding to a linear polarization at a certain angle $\psi$ with respect to the horizontal direction. We extract only a slight angle deviation of $\psi = -8.4°$ of the two equivalent linear polarization states, indicating correct operation of the PR. **d,** Measurement of the PER for both outputs showing very broadband operation over 350 nm with PER in excess of 11 dB.

For each of the output ports, we then extract the polarization extinction ratio (PER), which is here defined as the ratio of the maximum transmitted power at the target output polarization state to the minimum power at the antipodal point on the Poincaré sphere. For better reliability of the extracted results, we implemented a PER evaluation technique that considers all power levels recorded for the various input polarization states rather than just the maximum and the minimum power, see Supplementary Section S3 for details. The experiment was repeated for different wavelengths over a broad range from 1270 nm to 1620 nm, see Fig. S1d (which is identical to Fig. 4b of the main manuscript) for a plot of the extracted PER vs. wavelength. We find that the PER is better than 11 dB over the whole wavelength range, which was only limited by the tuning range of the underlying external-cavity lasers, (Ando AQ4321D, TUNICS T1005-HP, and Agilent 81600B). This is slightly worse than the performance expected by simulations, see Fig. 2c of the main manuscript. We also measured the insertion loss of the device using the polarization state of maximum transmission at each output port. At a wavelength of 1550 nm, we find losses of 4.4 dB and 3.8 dB for Output I and Output II, respectively. Note that these values include the loss of the PBS, of the subsequent PR, and of the adiabatic mode-field adapters at the input and the output of the device. Still, there is room for improvement considering the sub-2 dB losses expected from simulations of the PBS structure only, see Fig. 2c of the main manuscript. We attribute the deviations between measurements and simulations to imperfections of the 3D-printed structure, caused by limited resolution and shrinkage of the resist structure upon development. Exploiting super-resolution 3D-lithography inspired by the concept of stimulated-emission-depletion (STED)[3] microscopy might allow to better resolve fine details of the PBS structure and to further enhance the performance of the devices in the future.

## S2. Scattering parameters, Jones matrix, and PER

Figure 2c of the main manuscript gives the simulated PBS performance in terms of transmission, crosstalk, leakage, unfilterable crosstalk, and PER. The first four parameters are directly extracted from the corresponding elements of the simulated scattering matrix, as indicated in Fig. 2c and in the main text. For calculating the PER, we use the ratio of the squares of the singular values of the simulated Jones matrices of the PBS[4]. The Jones matrix associated with a certain output port describes the propagation of light from the PBS input port to this output port. In the following, the Jones matrix associated with output port $H$ is denoted as $\mathbf{T}_{\mathrm{PBS},H}$, while $\mathbf{T}_{\mathrm{PBS},V}$ refers to output port $V$. The Jones vector at the input port is $\mathbf{J}_I = \begin{bmatrix} E_H^{(I)} & E_V^{(I)} \end{bmatrix}^{\mathrm{T}}$, while the Jones vectors at the output ports $H$ and $V$ are $\mathbf{J}_H = \begin{bmatrix} E_H^{(H)} & E_V^{(H)} \end{bmatrix}^{\mathrm{T}}$ and $\mathbf{J}_V = \begin{bmatrix} E_H^{(V)} & E_V^{(V)} \end{bmatrix}^{\mathrm{T}}$. The Jones-matrix elements can be directly taken from the scattering matrix, such that the relations $\mathbf{J}_H = \mathbf{T}_{\mathrm{PBS},H}\,\mathbf{J}_I$ and $\mathbf{J}_V = \mathbf{T}_{\mathrm{PBS},V}\,\mathbf{J}_I$ between the Jones vectors at the input and at the output can be written as

$$\begin{bmatrix} E_H^{(H)} \\ E_V^{(H)} \end{bmatrix} = \begin{bmatrix} \underline{S}_{E_H^{(H)} E_H^{(I)}} & \underline{S}_{E_H^{(H)} E_V^{(I)}} \\ \underline{S}_{E_V^{(H)} E_H^{(I)}} & \underline{S}_{E_V^{(H)} E_V^{(I)}} \end{bmatrix} \begin{bmatrix} E_H^{(I)} \\ E_V^{(I)} \end{bmatrix}, \tag{S1}$$

$$\begin{bmatrix} E_H^{(V)} \\ E_V^{(V)} \end{bmatrix} = \begin{bmatrix} \underline{S}_{E_H^{(V)} E_H^{(I)}} & \underline{S}_{E_H^{(V)} E_V^{(I)}} \\ \underline{S}_{E_V^{(V)} E_H^{(I)}} & \underline{S}_{E_V^{(V)} E_V^{(I)}} \end{bmatrix} \begin{bmatrix} E_H^{(I)} \\ E_V^{(I)} \end{bmatrix}. \tag{S2}$$

The PER is then calculated as the ratio of the squares of the singular values $s_1$ and $s_2$ of the corresponding Jones matrices,[4]

$$\mathrm{PER} = \frac{s_1^2(\mathbf{T}_{\mathrm{PBS},H})}{s_2^2(\mathbf{T}_{\mathrm{PBS},H})} = \frac{s_1^2(\mathbf{T}_{\mathrm{PBS},V})}{s_2^2(\mathbf{T}_{\mathrm{PBS},V})}, \tag{S3}$$

where $s_1 \geq s_2$ without loss of generality.

## S3. PER extraction from the measurements

The PER of an optical device is generally defined as the ratio of maximum to minimum output power $P_{\mathrm{out}}$ that can be found when varying the input polarization over all possible states. In our experiments, the input polarization states were sampled randomly, and a straightforward way of calculating the PER is taking the ratio of the maximum to the minimum recorded output power. However, this approach takes into account only two measured power levels, which bears the risk that the result is subject to noise, which could lead to an overestimated PER. In addition, there is no guarantee that the sampled input states will fall close enough to the states of minimum and maximum transmitted power.

We therefore implemented a PER evaluation technique that considers all power levels recorded for the various input polarization states and relies on fitting a theoretical curve to the full set of measurement data. To explain this technique, we consider only one output port of the 3D-printed polarization-beam-splitter/polarization-rotator combination (PBS/PR) – the other output port can be treated in an analogous way. We represent four-dimensional normalized Stokes vectors $\mathbf{S} = \begin{bmatrix} 1 & S_1/S_0 & S_2/S_0 & S_3/S_0 \end{bmatrix}^{\mathrm{T}}$ by the corresponding three-dimensional Stokes vectors $\mathbf{s} = \begin{bmatrix} s_1 & s_2 & s_3 \end{bmatrix}^{\mathrm{T}}$, where $s_1 = S_1/S_0$, $s_2 = S_2/S_0$, and $s_3 = S_3/S_0$, that can be represented in the Cartesian coordinate system of the Poincaré sphere, see Section 14.5 of Ref.[5]. For simplicity, we further assume that the maximum power transmission for the considered port occurs for a perfectly horizontal ($x$-polarized) polarization at both the input and the output of the PBS/PR, characterized by three-dimensional Stokes vectors $\mathbf{s}_{\mathrm{in,pass}} = \mathbf{s}_{\mathrm{out,pass}} = \begin{bmatrix} 1 & 0 & 0 \end{bmatrix}^{\mathrm{T}}$. Note that the input port can only be accessed through an optical fiber that is connected to the polarization scrambler, and that the measurement of the power and the polarization state at the PBS/PR output requires a second optical fiber leading to the polarization analyzer, see Fig. S2a for a sketch of the experimental setup. In the following, we assume fully polarized light such that we can use either Stokes or Jones calculus, as appropriate. We describe the input fiber between the polarization scrambler and the PBS/PR by a Jones matrix $\mathbf{U}$, whereas the output fiber is described by a Jones matrix $\mathbf{V}$, see Fig. S2a. For a given polarization state with Jones vector $\mathbf{J}_{scr}$ emitted by the polarization scrambler, the Jones vector of the polarization state $\mathbf{J}_{an}$ received by the polarization analyzer can then be written as

$$\mathbf{J}_{an} = \mathbf{V} \, \mathbf{T}_{PBS} \, \mathbf{U} \, \mathbf{J}_{scr}, \tag{S4}$$

where $\mathbf{T}_{PBS}$ corresponds to the Jones matrix of the non-ideal PBS/PR, and where the Jones matrices $\mathbf{U}$ and $\mathbf{V}$ of the input fiber and the output fiber can be assumed to be unitary, see Fig. S2a. The light at the PBS/PR input can be expressed by an input Jones vector $\mathbf{J}_{in} = \mathbf{U} \, \mathbf{J}_{scr}$, which is characterized by an angle $\alpha_{in}$ that defines the ratio of the field amplitudes in the two polarizations and by a phase difference $\varphi$ between the $x$- and the $y$-component,

$$\mathbf{J}_{in} = \underline{E}_{in} \begin{bmatrix} \cos(\alpha_{in})e^{-j\varphi/2} \\ \sin(\alpha_{in})e^{+j\varphi/2} \end{bmatrix}. \tag{S5}$$

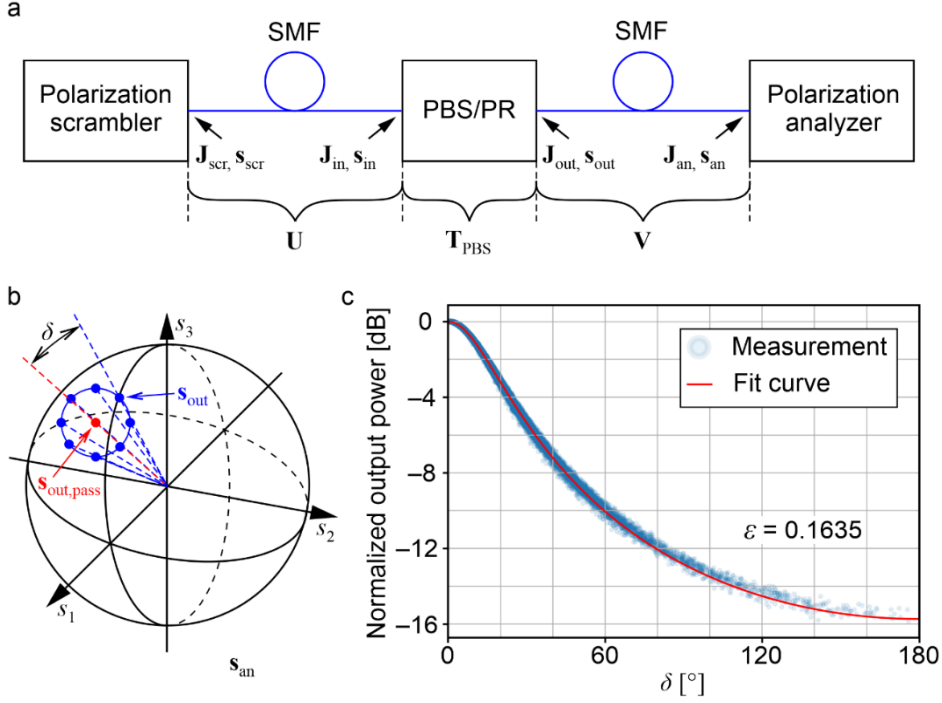

**Figure S2: Measurement and evaluation of the polarization extinction ratio (PER) of the 3D-printed polarization-beam-splitter/polarization-rotator (PBS/PR) combination, taking into account the full set of measured output powers and polarization states. a,** Experimental setup: Light emitted from a polarization scrambler is fed to the 3D-printed PBS/PR through a standard single-mode fiber (SMF), and the output power and the output polarization state are measured by a polarization analyzer, which is connected to the PBS/PR by a second SMF. $\mathbf{J}_{scr}$, $\mathbf{J}_{in}$, $\mathbf{J}_{out}$, and $\mathbf{J}_{an}$ denote the Jones vectors at the output of the polarization scrambler, the input and the output of the PBS/PR, and at the input of the polarization analyzer, respectively, and $\mathbf{s}_{scr}$, $\mathbf{s}_{in}$, $\mathbf{s}_{out}$, and $\mathbf{s}_{an}$ are the corresponding Stokes vectors. Note that we represent four-dimensional normalized Stokes vectors $[1 \quad S_1/S_0 \quad S_2/S_0 \quad S_3/S_0]^T$ by the corresponding three-dimensional Stokes vectors $\mathbf{s} = [s_1 \quad s_2 \quad s_3]^T$, where $s_1 = S_1/S_0$, $s_2 = S_2/S_0$, and $s_3 = S_3/S_0$, that can be represented in the Cartesian coordinate system of the Poincaré sphere. The non-ideal PBS/PR is modeled by a Jones matrix $\mathbf{T}_{PBS}$, while the two SMF at the input and the output side of the PBS/PR are represented by two unitary Jones matrices $\mathbf{U}$ and $\mathbf{V}$, respectively. **b,** Illustration of the three-dimensional vectors $\mathbf{s}_{out}$ recorded by the polarization analyzer. Since we assume fully polarized light, all vectors are on the surface of the Poincaré sphere. The output power should be the same for all polarization states that are located on a circle, which is centered about the state of maximum transmission. The radius of this circle is quantified by the opening angle $\delta$ of the associated cone, which can be directly connected to the normalized output power. **c,** Normalized output power $P_{out}/P_{in}$ vs. angle $\delta$, as recorded for the data point for Output 1 at a wavelength of 1460 nm, see Fig. S1d. By fitting a model function (red) to the measurement data (blue), we extract a polarization leakage magnitude of $\varepsilon = 0.1635$, corresponding to a PER of 15.7 dB.

In this relation, $\underline{E}_{in}$ denotes the electric field that is associated with the signal at the input of the 3D-printed PBS/PR – the corresponding power is denoted by $P_{in} \sim |\underline{E}_{in}|^2$. For the PBS/PR, we assume a simplified Jones matrix $\mathbf{T}_{PBS}$ that corresponds to that of a non-ideal linear polarizer oriented along the $x$-direction,

$$\mathbf{T}_{\mathrm{PBS}} = \begin{bmatrix} 1 & 0 \\ 0 & \varepsilon \end{bmatrix}, \tag{S6}$$

where $\varepsilon$, $0 \leq \varepsilon \leq 1$, is the magnitude of the polarization leakage. The corresponding PER is then found as the ratio of the squares of the singular values of $\mathbf{T}_{\mathrm{PBS}}$[4]

$$\mathrm{PER} = \frac{1}{\varepsilon^2}. \tag{S7}$$

Note that the model for the Jones matrix according to Eq. (S6) represents an approximation: The Jones matrices $\mathbf{T}_{\mathrm{PBS},H}$ and $\mathbf{T}_{\mathrm{PBS},V}$ that are obtained from our simulations, Eqs. (S1) and (S2), do have non-zero off-diagonal elements and are generally not Hermitian. As a consequence, transformation into a diagonal matrix as assumed in Eq. (S6) is not generally possible. Still, the magnitudes of the off-diagonal elements are small such that the associated error should not be severe, see discussion below.

Using the Jones-matrix model according to Eq. (S6), the relation between a given polarization state, $\mathbf{J}_{\mathrm{in}}$ at the input of the PBS/PR and the corresponding output state $\mathbf{J}_{\mathrm{out}}$ can be written as

$$\mathbf{J}_{\mathrm{out}} = \mathbf{T}_{\mathrm{PBS}} \cdot \mathbf{J}_{\mathrm{in}} = \underline{E}_{\mathrm{in}} \begin{bmatrix} \cos(\alpha_{\mathrm{in}}) e^{-\mathrm{j}\varphi/2} \\ \varepsilon \sin(\alpha_{\mathrm{in}}) e^{+\mathrm{j}\varphi/2} \end{bmatrix}. \tag{S8}$$

We can now express the ratio of the power $P_{\mathrm{out}}$ at the output of the PBS/PR to the input power $P_{\mathrm{in}}$ in terms of the magnitude of the polarization leakage $\varepsilon$ and the angle $\alpha_{\mathrm{in}}$,

$$\frac{P_{\mathrm{out}}}{P_{\mathrm{in}}} = \frac{|\mathbf{J}_{\mathrm{out}}|^2}{|\mathbf{J}_{\mathrm{in}}|^2} = \cos^2(\alpha_{\mathrm{in}}) + \varepsilon^2 \sin^2(\alpha_{\mathrm{in}}). \tag{S9}$$

Note that the ratio in Eq. (S9) does not depend on the phase difference $\varphi$.

When evaluating the measurement, we face the problem that the angle $\alpha_{\mathrm{in}}$ and thus the expression for the power transmission according to Eq. (S9) are related to the Jones vector at the output of the PBS/PR, which cannot be accessed in the measurement. To establish a relationship to the known polarization state $\mathbf{J}_{\mathrm{an}}$ at the input of the polarization analyzer, we proceed in two steps. First, we switch to Stokes space, and we find a relationship that connects the angle $\alpha_{\mathrm{in}}$ and the magnitude of the polarization leakage $\varepsilon$ in Eq. (S8) to the angle $\delta$ between the actual three-dimensional Stokes vector $\mathbf{s}_{\mathrm{out}}$ at the PBS/PR output and the three-dimensional Stokes vector $\mathbf{s}_{\mathrm{out,pass}} = \begin{bmatrix} 1 & 0 & 0 \end{bmatrix}^{\mathrm{T}}$ that corresponds to maximum transmission. To this end, we first calculate $\mathbf{s}_{\mathrm{out}} = \begin{bmatrix} s_{\mathrm{out},1} & s_{\mathrm{out},2} & s_{\mathrm{out},3} \end{bmatrix}^{\mathrm{T}}$ from the components of vector $\mathbf{J}_{\mathrm{out}}$ using Eqs. (6.1-9a)–(6.1-9d) in Ref. [6]. The angle $\delta \in [0, \pi]$ between the measured three-dimensional Stokes vector $\mathbf{s}_{\mathrm{out}}$ and the three-dimensional Stokes vector $\mathbf{s}_{\mathrm{out,pass}} = \begin{bmatrix} 1 & 0 & 0 \end{bmatrix}^{\mathrm{T}}$ of maximum transmission can then be calculated as

$$\cos(\delta) = \mathbf{s}_{\mathrm{out}} \cdot \mathbf{s}_{\mathrm{out,pass}} = s_{\mathrm{out},1} = \frac{\cos^2(\alpha_{\mathrm{in}}) - \varepsilon^2 \sin^2(\alpha_{\mathrm{in}})}{\cos^2(\alpha_{\mathrm{in}}) + \varepsilon^2 \sin^2(\alpha_{\mathrm{in}})}, \tag{S10}$$

which can be simplified to

$$\tan\left(\frac{\delta}{2}\right) = \varepsilon \tan(\alpha_{\mathrm{in}}). \tag{S11}$$

In a second step, we then account for the propagation of the signal from the PBS/PR output to the polarization analyzer. To this end, we exploit the fact that the corresponding Jones vectors $\mathbf{J}_{\mathrm{out}}$ and $\mathbf{J}_{\mathrm{an}}$ are related by a unitary transformation that is described by the Jones matrix $\mathbf{V}$ of the output fiber. In the Cartesian coordinate system of the Poincaré sphere, this transformation simply corresponds to a rotation about the origin, which leaves the relative angle $\delta$ between the measured vectors $\mathbf{s}_{\mathrm{out}}$ and $\mathbf{s}_{\mathrm{out,pass}}$ unchanged. In other words: For a given polarization leakage magnitude $\varepsilon$, the output power $P_{\mathrm{out}}$ should be the same for all polarization states that are located on a circle on the surface of the Poincaré sphere which is centered about $\mathbf{s}_{\mathrm{out,pass}}$, see Fig. S2b for an illustration. We may thus extract this angle directly from the polarization states recorded at the polarization analyzer, where $\mathbf{s}_{\mathrm{out,pass}}$ corresponds to polarization state for which the highest output power was measured. We then use Eq. (S11) with $\varepsilon$ as a parameter to extract $\alpha_{\mathrm{in}}$ and predict the dependence of the power $P_{\mathrm{out}}$ on $\delta$ via Eq. (S9), assuming constant $P_{in}$. We finally vary the magnitude of the polarization leakage $\varepsilon$ to find best coincidence between the measured $\delta$-dependence of $P_{out}$ and the associated model prediction, see Fig. S2c. Equation (S7) then allows us to calculate the PER for this value of $\varepsilon$.

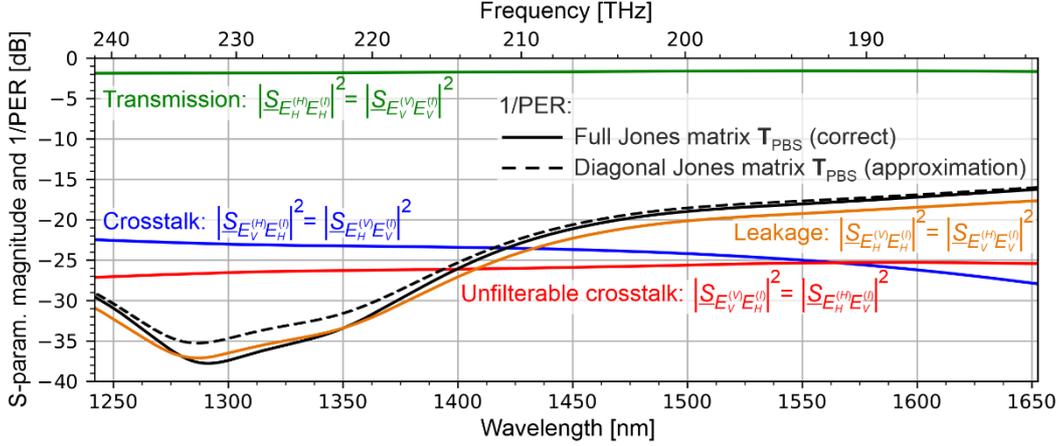

**Figure S3:** Comparison of PER extracted from the simulated Jones matrices without any off-diagonal elements according to the simplified model in Eq. (S6) (dashed black lines) and the PER extracted from the full Jones matrix (solid black line). The device is the same as the one described by Fig. 2c of the main manuscript. For better comparison, we also give the transmission, the crosstalk, the leakage, and the unfilterable crosstalk of the device – they are identical to the curves in Fig. 2c of the main manuscript.

We show the results of this technique in Fig. S2c for the highest PER that we measured during our wavelength sweep, i.e., for Output 1 at a wavelength of 1460 nm, see Fig. S1d. From the least-squares model fit shown in Fig. S2c, we estimate a field leakage $\varepsilon$ of 0.1635, corresponding to a PER of 15.7 dB. To check the validity of the approach, we also extract the PER by simply taking the ratio of the maximum and the minimum transmitted power, which leads to value of 16.1 dB. This confirms the validity of our approach, in particular with respect to the simplified model for the Jones matrix according to Eq. (S6). The result is also in line with the expectation that the PER extracted from the ratio of the maximum and the minimum transmitted power might be slightly overestimate due to measurement noise. We further checked the impact of neglecting the off-diagonal Jones-matrix elements in Eq. (S6) by simulations. To this end, we omit the elements $\underline{S}_{E_H^{(H)} E_V^{(I)}}$, $\underline{S}_{E_V^{(H)} E_H^{(I)}}$, $\underline{S}_{E_H^{(V)} E_V^{(I)}}$, and $\underline{S}_{E_V^{(V)} E_H^{(I)}}$ of the simulated Jones matrices according to Eqs. (S1) and (S2) and then extract the PER via Eq. (S7). The resulting PER is then compared to the one extracted from the singular values of the full Jones matrices, see Fig. S3. We find that omitting the off-diagonal Jones-matrix elements leads to a slight reduction of the extracted PER, and we conclude that the simplification related to Eq. (S6) does not bear the risk to overestimate the PER in our experiments.

## S4. Data transmission experiment

The setup used for data transmission experiment is depicted in Fig. S4. To generate a 16QAM data stream at a symbol rate of 80 GBd, a dual-polarization (DP) IQ modulator is driven by a high-speed arbitrary waveform generator (AWG, Keysight M8194A 120 GSa/s) using random bit sequences with different seeds for each polarization. The optical carrier at a wavelength of 1550 nm is provided by an external-cavity laser (ECL, Keysight N7714A, emission frequency $f_c$ in Inset 1 of Fig. S4). Root-raised-cosine pulse shaping at a roll-off factor of $\beta = 0.1$ is used for good spectral efficiency. At a BER of $1.25 \times 10^{-2}$, which corresponds to the threshold of forward error correction with 15 % coding overhead, see Table 7.5 in Ref. [7], our transmission setup exhibits an OSNR penalty of approximately 3 dB with respect to an ideal transmission system, see Fig. 4d of the main manuscript. This is in accordance with values in literature for similar modulation formats and symbol rates[8].

For the OSNR sweep at the receiver, band-limited amplified stimulated emission (ASE) noise is generated by a dedicated ASE noise source (Orion Laser Technology ASE-C/CL) and added to the optical signal (ASE noise loading). The noise-loaded signal is then amplified by an EDFA, filtered by a bandpass filter (BPF, full width at half maximum 1 nm) to suppress out-of-band amplified ASE noise, and sent to the PBS, which may be either a 3D-printed PBS/PR assembly or a commercial fiber-based PBS that we use as a reference. After the PBS, each polarization is detected using a coherent heterodyne scheme, where the local oscillator laser (LO, Keysight N7714A, emission frequency $f_{LO}$ in Inset 1 of Fig. S4) is tuned to the edge of the signal spectrum. Two balanced photodetectors (BPD, Finisar BPDV2150RQ) are used to suppress both signal-signal and LO-LO mixing products. The outputs of the BPD are digitized by a 256 GSa/s real-time oscilloscope (Keysight UXR1004A) and recorded for offline digital signal processing (DSP). In a first DSP step, the signals are made analytic and are shifted in frequency by the difference between the carrier and the LO. After timing recovery, a $2 \times 2$ MIMO equalizer is used for polarization de-multiplexing, and afterward the carrier recovery is performed. The MIMO equalizer is an adaptive equalizer, whose coefficients are updated according to the radius directed equalization (RDE)[9]. Finally, the signals go through a least-mean-square equalizer before being decoded. To benchmark the performance of the PBS/PR assembly, the experiment is also performed with a commercially available PBS (AFW Technologies, POBS-15). Since the commercially available PBS exhibits less insertion loss than the PBS/PR-fanout assembly, we adjust the amplification of the preamplifier to obtain equal powers at the inputs of the BPD in both cases.

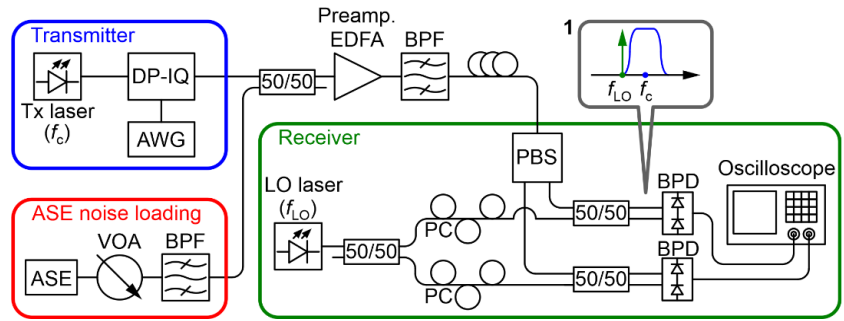

**Figure S4: Experimental setup for the data-transmission demonstration:** An optical carrier at $\lambda = 1550$ nm (frequency $f_c$) is modulated by a dual polarization IQ (DP-IQ) modulator that is driven by an arbitrary waveform generator (AWG) to generate a 16QAM PDM signal at 80 GBd. The band-limited amplified-spontaneous-emission (ASE) source generates noise, whose power is varied by a variable optical attenuator (VOA), and added to the 16QAM signal. This noise-loaded data signal is amplified by an EDFA, filtered by a bandpass filter (BPF), and guided to the PBS input in the receiver block. A local oscillator (LO) signal (frequency $f_{LO}$) is split, and the two split signals are sent through a pair of polarization controllers (PC) and superimposed with the two output signals of the PBS in a pair of balanced photodetectors (BPD). The electrical output signals are detected by a high-speed oscilloscope. Inset 1 illustrates the spectrum at the BPD inputs, with the LO tone tuned to the edge of the signal for heterodyne detection.